# Quantum Transport and Band Structure Evolution under High Magnetic Field in Few-Layer Tellurene

Gang Qiu[1,2], Yixiu Wang[3], Yifan Nie[4], Yongping Zheng[4], Kyeongjae Cho[4], Wenzhuo Wu[2,3],

Peide D. Ye[1,2]⋆

[1]School of Electrical and Computer Engineering, Purdue University, West Lafayette, Indiana 47907, USA

[2]Birck Nanotechnology Center, Purdue University, West Lafayette, Indiana 47907, USA

[3]School of Industrial Engineering, Purdue University, West Lafayette, Indiana 47907, USA

[4]Department of Materials Science and Engineering, the University of Texas at Dallas, Richardson, Texas 75080, United States

⋆e-mail: Correspondence and requests for materials should be addressed to P. D. Y. (yep@purdue.edu)

**Quantum Hall effect (QHE) is a macroscopic manifestation of quantized states which only occurs in confined two-dimensional electron gas (2DEG) systems. Experimentally, QHE is hosted in high mobility 2DEG with large external magnetic field at low temperature. Two-dimensional van der Waals materials, such as graphene and black phosphorus, are considered interesting material systems to study quantum transport, because it could unveil unique host material properties due to its easy accessibility of monolayer or few-layer thin films at 2D quantum limit. Here for the first time, we report direct observation of QHE in a novel low-dimensional material system – tellurene. High-quality 2D tellurene thin films were acquired from recently reported hydrothermal method with high hole mobility of nearly 3,000 $cm^2$/Vs at low temperatures, which allows the observation of well-developed Shubnikov-de-Haas (SdH) oscillations and QHE. A four-fold degeneracy of Landau levels in SdH oscillations and QHE was revealed. Quantum oscillations were investigated under different gate biases, tilted magnetic fields and various temperatures, and the results manifest the inherent information of the electronic structure of Te. Anomalies in both temperature-dependent oscillation amplitudes and transport characteristics were observed which are ascribed to the interplay between Zeeman effect and spin-orbit coupling as depicted by the density functional theory (DFT) calculations.**

Two-dimensional electron gas (2DEGs) is an important subject in condensed-matter physics for being a host to many exotic and conceptual physical phenomena. Quantum Hall Effect (QHE), which refers to the effect of the transverse resistance being quantized into integer fractions of quantity $h/e^2$ in Hall measurement, is such a phenomenon that can only be observed in high-mobility 2DEGs. Traditionally, the confinement of electron motions within a two-dimensional plane is realized by quantum wells formed in inversion channel of silicon metal-oxide-semiconductor field-effect transistors (MOSFETs)[1] or AlGaAs/GaAs heterojunctions[2]. In the last decade, the emergence of 2D materials offers an alternative approach to achieving high mobility 2DEGs in atomically thin layers. 2D materials are formed by an assembly of atomic layers bonded by weak van der Waals forces. The crystal structure of 2D materials yields a dangling-bond-free surface and consequently the 2D material systems are less susceptible to mobility degradation with thickness shrinking down to a few nanometers than bulk materials. QHE has been observed only in a few high-mobility 2D material systems such as graphene[3–5], InSe[6] and encapsulated few-layer phosphorene[7–9] among thousands of predicted 2D materials[10]. Shubnikov-de-Haas (SdH) oscillations were observed in several other 2D materials such as $WSe_2$[11,12], $MoS_2$[13–15], $ZrTe_5$[16–19], and $Bi_2O_2Se$[20]. Here for the first time, we report pronounced SdH oscillations and QHE in a novel 2D material: high-mobility air-stable few-layer Te films, coined as tellurene.

Tellurium (Te) is a group VI narrow bandgap p-type semiconductor with unique one-dimensional van der Waals crystal structure. The atoms are covalently bonded into skew-symmetric helical chains, and the neighboring chains are bonded by van der Waals forces in the other two dimensions to form a trigonal crystal structure[21–23]. With proper crystal growth techniques, tellurium can also be arranged in a plane to form 2D morphology with bond-free surfaces (see Figure 1a). Bulk Te has a direct bandgap of ~0.35 eV[24] at the corner of Brillouin zone H (and H') point, as shown in Figure 1b. The broken spatial-inversion-symmetry of Te lattice structure gives rise to strong Rashba-like spin-orbit interaction[24], therefore a camelback-like electronic structure at the edge of the valence band[22,25,26] were predicted by the density functional theory (DFT) calculations (right side of Figure 1b, the DFT computational details and complete band structure are presented in Supplementary Note 3 and 4). Weyl nodes were also theoretically predicted[24] and experimentally observed deeply embedded in the valence band of Te[27]. The earliest study on surface quantum states in bulk Te dates back to the early 1970's by von Klitzing, et al.[28–30], who later discovered QHE in silicon MOSFETs in 1980[1]. However, to our best knowledge, the evidence of QHE in Te is so far still missing, and even the reported quantum oscillations were so weak that it could only be seen in the second derivative $d^2R_{xx}/dB^2$ [28,29]. Since Te has superior mobility compared to silicon, it is intriguing why quantum phenomena are so weak in Te. We postulate that this is due to the weak quantization of Te surface states in the wide triangular potential well. The characteristic width of the surface potential well λ is proportional to $\sqrt{\epsilon_s}$ where $\epsilon_s$ is the permittivity of the semiconductor. The large permittivity of tellurium (~30) only induces a relatively small electrical field and wide potential well in tellurium surface. Therefore, the carriers are not tightly restrained near the surface to meet the rigid criteria of 2D confinement. In this work, we adopted a different strategy to impose stronger quantum confinement by employing 2D Te films. In general, the most widely used methods to acquire 2D thin films from van der Waals materials are mechanical exfoliation and chemical vapor deposition (CVD). Significant efforts have also been made to grow tellurium nanostructures, however these methods yield either quasi-one-dimensional morphology[31–36] or small 2D flakes[37] that are unsuitable for studying magneto-transport. It is also very difficult to study transport properties of monolayer tellurene grown on conducting graphene surface by molecular beam epitaxy[38,39] since it requires monolayer transfer and process[40–42]. Recently a liquid-based growth method[43] was proposed to obtain large-scale 2D tellurene films with lateral sizes over 50 um ×100 um and thicknesses ranging from tens of nanometers down to few layers. These nano-films were verified with high crystal quality by crystallographic and spectroscopic techniques and offered a new route to explore the potential for electronic device applications[43] and magneto-transport studies[44]. To study quantum transport, flakes with thickness around 8-10 nm are desired because (1) high crystal quality is still preserved with hole mobility nearly 3,000 cm$^2$/Vs at liquid Helium temperature; (2) the physical confinement in the out-of-plane direction is strong enough to impose sufficient 2D confinement.

The liquid-solution based Te flake synthesis method was described in ref. 42 (also see Supplementary Note 1). As-grown tellurene flakes were dispensed onto heavily doped p-type Silicon wafer with 300 nm SiO$_2$ capping layer which serves as a universal back gate to modulate the Fermi level of Te samples. Standard six-terminal Hall-bar device was defined by electron beam lithography and metallization was carried out by electron beam evaporator. The hydrothermal growth method usually yields rectangular or trapezoidal flakes (see inset of Figure 1c) where the helical axis of the crystal is aligned with the long edge of the flakes, which facilitates our fabrication and measurement. This property was confirmed by detailed physical characterization such as transmission electron microscopy (TEM) and angle-resolved Raman spectroscopy[43,44]. In order to verify the anisotropic transport, we used dry etching technique to trim the flakes into two identical rectangles but in different crystal orientations: one along atomic chain direction (c-axis) and the other one along van der Waals force direction (a-axis), and fabricated Hall bar structures accordingly (device fabrication process and characterization see Supplementary Note 2). The optical image of a typical pair of devices fabricated on a 10-nm-thick flake was shown in Figure 1c. We determined that the helical chain direction exhibits ~30% higher Hall mobility (more discussion see Supplementary Note 5 and 6). This ratio is close to the field-effect mobility anisotropic ratio measured at room temperature[43]. It was also noticed that the SdH oscillations and QHE features were more distinguishable along a-axis. Since the SdH oscillations are related with the in-plane cyclotron motions and QHE plateaus occur at metrological standard value of resistance irrelevant of device geometry, the data we presented in the latter part of the paper is acquired from devices along a-axis simply for its stronger oscillation signal and soundness of data interpretation.

Figure 1d shows the typical longitudinal resistance ($R_{xx}$) and transverse resistance ($R_{xy}$) of a 10-nm-thick sample as a function of out-of-plane magnetic field measured at 300 mK with -85V gate bias. From the slope of $R_{xy}$ in the small B-field regime, the sheet carrier density of 8.7×10$^{12}$ cm$^{-2}$ (positive slope indicates hole is dominant carrier type) was extracted from the equation $n_{2D} = 1/eR_H$, where $R_H$ is the Hall coefficient, with corresponding Hall mobility of 2,640 cm$^2$/Vs derived from $\mu_{Hall} = \frac{L}{W}\frac{1}{R_{xx}n_{2D}e}$. Well-developed SdH oscillations were observed in longitudinal resistance, and the onset occurs at about 5 Tesla, from which we can estimate the quantum mobility to be ~ 2,000 cm$^2$/Vs from the criteria of SdH oscillation onset: $\omega_c \tau > 1$ (cyclotron frequency $\omega_c = eB/m^*$, $\tau$ is the relaxation time and $m^*$ is the hole effective mass in Te films), close to the mobility value obtained from Hall measurement. At large B-fields the transverse resistance shows a series of quantum Hall plateaus at the values of integer fraction of resistance quanta h/e$^2$, although the plateaus are not completely flat due to the broadening effect. The filling factors were calculated by normalizing R$_{xy}$ with metrological standard value of h/e$^2$. With the capacity of our equipment of 18 Tesla magnetic field, the lowest filling factor ν we can observe is 20. An interval of 4 was found between the filling factors of neighboring plateaus, indicating a four-fold Landau level degeneracy. The four-fold degeneracy can be factored into two parts: (1) the spin degeneracy factor g$_s$=2; and (2) valley degeneracy factor g$_v$=2 originating from H and H' point in Brillouin zone, similar to the case in graphene (K and K' points). Our tellurene samples show much better-resolved SdH oscillations than previously works on surface states of bulk Te, due to the high quality of 2D tellurene films and enhanced quantum confinement. We also found that QHE and SdH oscillations become weaker and eventually undetectable in thicker flakes (see Supplementary Note 7).

We then applied different gate bias to tune the carrier density and studied the gate-dependence of oscillation frequencies. The oscillation amplitudes were extracted by subtracting a smooth background from original data of R$_{xx}$ and were plotted versus 1/B in Figure 2a. Strong oscillations with constant frequency Δ (1/B) are well-developed with a negative gate bias over -40 V. However, at lower (absolute value of) gate bias, no pronounced oscillations can be observed, because of the lower mobility at subthreshold regime where the carrier concentration is low. It indicates the impurity scattering is still dominant like most of the 2D materials and there is still large room to enhance transport properties through methods such as hexagonal boron nitride (h-BN) capping. Landau fan diagram was constructed by plotting the Landau level index versus the value of 1/B where the minimum of R$_{xx}$ occurs. All the data points fall onto a line that can be extrapolated towards zero intercept (fitting error within ± 0.15), indicating the standard phase of oscillations.

The oscillation frequency $B_F$ are derived from fast Fourier transform, from which 2D carrier density $n_s$ can be estimated by substituting the frequency into the equation of frequency-density relationship: $n_s = g_s g_v e B_F/h$, where $g_s$ and $g_v$ are spin degeneracy and valley degeneracy respectively. We also compared the oscillation carrier density (blue spheres) and Hall density (red pentagrams) in Figure 2c. The orange dash line is calculated from the gate capacitance model $n_{2D} = \frac{1}{e} C_g (V_g - V_{th})$. The Hall densities, oscillation densities and densities calculated from capacitance model match quite well within a reasonable margin.

To better understand the two-dimensionality of Fermi surface in tellurene, we carried out angle-resolved SdH oscillation measurements in tilted magnetic fields. The experiment configuration is illustrated in Figure 3a. The sample was rotated along a-axis by angle α, and the corresponding 2D Fermi surface is associated with SdH oscillation frequency through the equation: $B_F(\alpha) = h S_F(\alpha)/4\pi^2 e$. The oscillation amplitudes with different angles were plotted in a color map in Figure 3b. A shift of oscillation features towards higher magnetic field was captured as the tilted angle increases. For clarity, the angle-resolved SdH oscillation amplitudes were plotted against perpendicular magnetic field component $B_\perp = B \cdot \cos(\alpha)$ in Figure 3c, where the minima of oscillations occur at the same value. The oscillations are too vague to be distinguished for α > 35º, and in 0º < α < 32º regime, the oscillation features also diminishes as we increase the angle (see Figure 3d, right axis). This is because in 2D films with thickness comparable to or smaller than the radius of cyclotron orbit of n-*th* Landau level $\sqrt{n} l_B$ (here $l_B = \sqrt{\hbar/eB}$ is the magnetic length which is 6.0 nm at 18 T), the cyclotrons cannot perform a complete circular motion under an oblique magnetic field with a large tilted angle, and hence the oscillation amplitude is reduced. By performing fast Fourier transform, we found that the oscillation frequency $B_F(\alpha)$, or the projection of Fermi surface $S_F(\alpha)$, is linear dependent on the factor 1/cos (α), another signature of two dimensionality of carrier motions.

Finally, we studied the temperature-dependence of SdH oscillations and the evidence of the band structure evolution with large magnetic field originated from the interplay of Zeeman splitting and spin-orbit coupling. Temperature-dependent SdH oscillation amplitudes were first investigated as a common practice to extract effective mass. As seen in Figure 4a, the period and the phase of the oscillations are independent of temperature, and the amplitudes exhibit a diminishing trend with rising temperature in general. For temperature higher than 5 K, the amplitudes decrease monotonically as we expected and can be fitted with the classical Lifshitz-Kosevich equation: $\Delta R_{xx} \propto \frac{2\pi^2 k_B m^*/\hbar eB}{\sinh(2\pi^2 k_B m^*/\hbar eB)}$. The fitted curves are plotted as solid lines in Figure 4c, and the effective mass m* can be extracted to be ~ 0.26 $m_0$ (here $m_0$ is the free electron mass) correspondingly, which is very close to our DFT calculation. However, by carefully examining the amplitude at each oscillation peak (zoomed in features are plotted in Figure 4b), we found that at temperature lower than 5 K the data points do not coincide with fitted curves, as shown in Figure 4b and 4c, and the discrepancy is more significant at larger B field. The deviation of oscillation amplitudes is abnormal, suggesting the average effective mass is heavier under the conditions of large magnetic fields and low temperatures. This can be tentatively explained by the band structure reshaping with the existence of external magnetic field. An insightful picture of the band structure evolution is given by DFT calculations under magnetic field perpendicular to the helical axis, with spin-orbit coupling taken into consideration, as shown in Figure 4d-e (The details of the DFT calculation is discussed in the Supplementary Information Note 3). The broken spatial-inversion-symmetry of Te crystal introduces Rashba-like spin-orbit coupling (SOC), which causes a lateral shift of spin-up and spin-down subbands in different directions along $k_z$ axis in band dispersion diagram and forms a camelback structure as shown in Figure 4d. With the presence of external B field, the degenerate energy levels for spin-up and spin-down states will also be lifted by Zeeman splitting energy: $E_z = g\mu_B B$, where $g$ is the g-factor and $\mu_B$ is the Bohr magneton. The interplay of SOC and Zeeman splitting introduces band evolution as shown in Figure 4e. If we take a cross-section of band dispersion along the camelback direction ($k_z$ direction), we can clearly see that the two degenerate valleys will be split into one heavy hole pocket and one light hole pocket with increasing magnetic field (Figure 4f). Noted that the oscillation features were captured under large negative gate bias, which is equivalent to pulling the Fermi level slightly inside the valence band, as shown in the dashed line in Figure 5c. At sufficiently low temperature condition, the Boltzmann distribution ensures that the states responsible for transport will be confined within a very narrow range (several kT) near Fermi level, and the carriers in the right branch of Figure 4f will dominate the transport behavior, which exhibits larger effective mass. As the temperature rises, such effect will be mitigated by temperature broadening as the fine electrical structure becomes undetectable and the carriers in both pockets will participate in transport, hence the average effective mass will fall back to the classical value. The DFT calculation suggests that as we increase the external B-field to 18 T, the effective mass of the light hole pocket slightly drops from 0.26 to 0.19 $m_0$, whereas the effective mass in the heavy hole pocket increases drastically to about 9.8 $m_0$ (see inset of Figure 4f), which qualitatively matches with our experiment results.

Another evidence related to this effect is the anomalous transfer characteristics of $R_{xx}$ versus gate bias under large magnetic field. For normal 2DEG systems, in the linear regime, the channel resistance should drop inversely as we raise overdrive voltage $|V_g-V_{th}|$. Here, $V_{th}$ is the threshold voltage from which the channel is turned on. However, in tellurene films under large external magnetic fields and at low temperatures, as we sweep the gate voltage in reverse direction, the sample undergoes a sudden rise in resistance at around -20 V before it continues to drop as shown in the top curve in Figure 5a. We further verified that this anomalous increase in resistance has a strong dependence on the magnetic field (see Figure 5a). Therefore we believe it is also related to the aforementioned Zeeman effect induced band structure evolution. We denote three representative situations (1), (2) and (3) as we lower the Fermi level in Figure 5c, corresponding to spots (1), (2) and (3) in the transfer curves in Figure 5a. At positions (1) and (3), the transport behavior can be explained with classical semiconductor picture. At position (2) however, the carriers in the heavy pocket start to dominate the transport and the resistance will increase because of larger average effective mass. This anomalous resistance bump will also vanish at higher temperature (shown in Figure 5b) as the temperature broadens the Boltzmann distribution, making it difficult to distinguish such fine structures in the electronic band.

In conclusion, we realized strong 2D confinement in a novel van der Waals 2D material tellurene by taking advantage of recently developed liquid-phase grown high-mobility air-stable tellurene ultrathin films. QHE was observed on Te for the first time with four-fold degeneracy. Well-developed Shubnikov-de-Haas oscillations were thoroughly investigated under different gate bias, temperatures, and tilted magnetic fields. Anomaly in cyclotron effective mass and transfer characteristics at large magnetic fields and low temperatures were also observed, which is attributed to the interplay of Zeeman splitting and strong Rashba-like spin-orbit coupling in Te. Our work revealed the intriguing electronic structure of Te through quantum transport and demonstrated high quality of the novel 2D tellurene material system which is suitable to even explore exotic topological phenomena in Te.

Figures

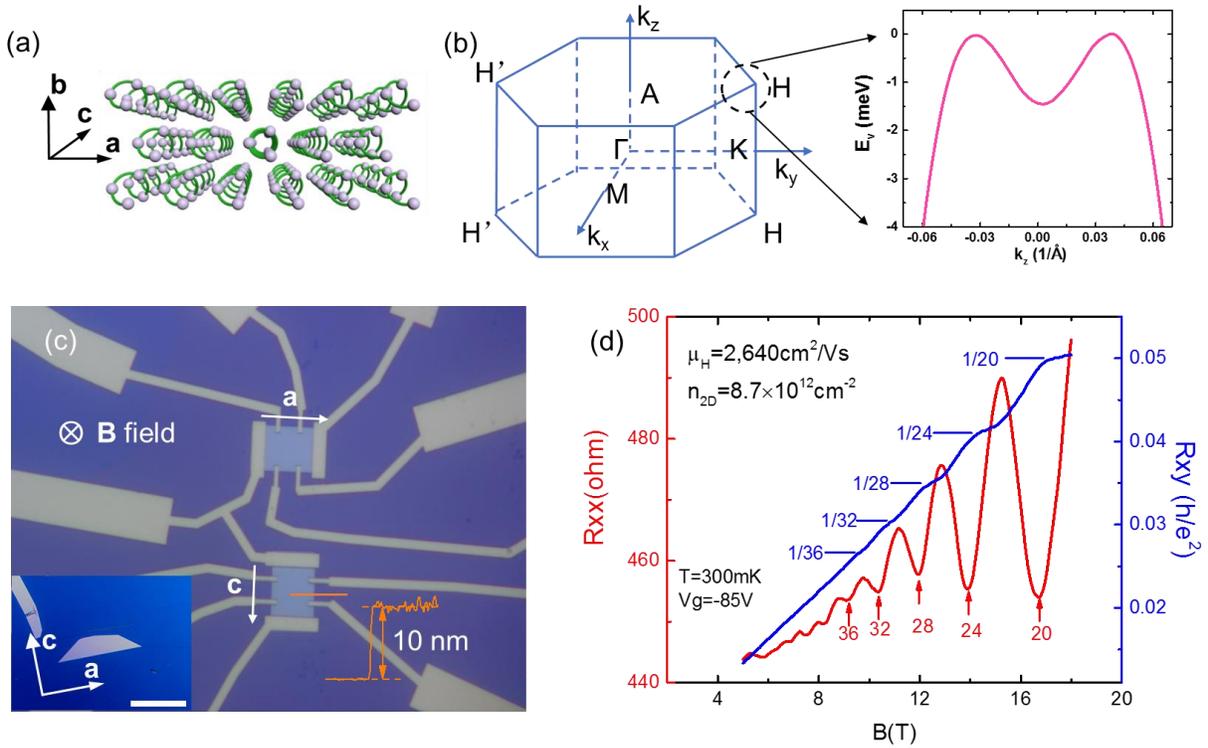

**Figure 1 (a)** Crystal structure of one-dimensional van der Waals material tellurium. The collection of helical atom chains can be arranged into 2D films termed as tellurene. **(b) Left**: Brillouin zone of Te in momentum space. **Right**: zoom-in features at H point of Brillouin zone where the valence band maxima reside. **(c)** An optical image of Hall-bar devices along two different crystal orientations fabricated from the same flake. **Inset**: as-grown flakes are usually in rectangular or trapezoidal shapes where the 1D direction (c-axis) is along the longer edge of the flake. The scale bar is 50 µm. **(d)** QHE and SdH oscillations measured from a 10-nm-thick flake.

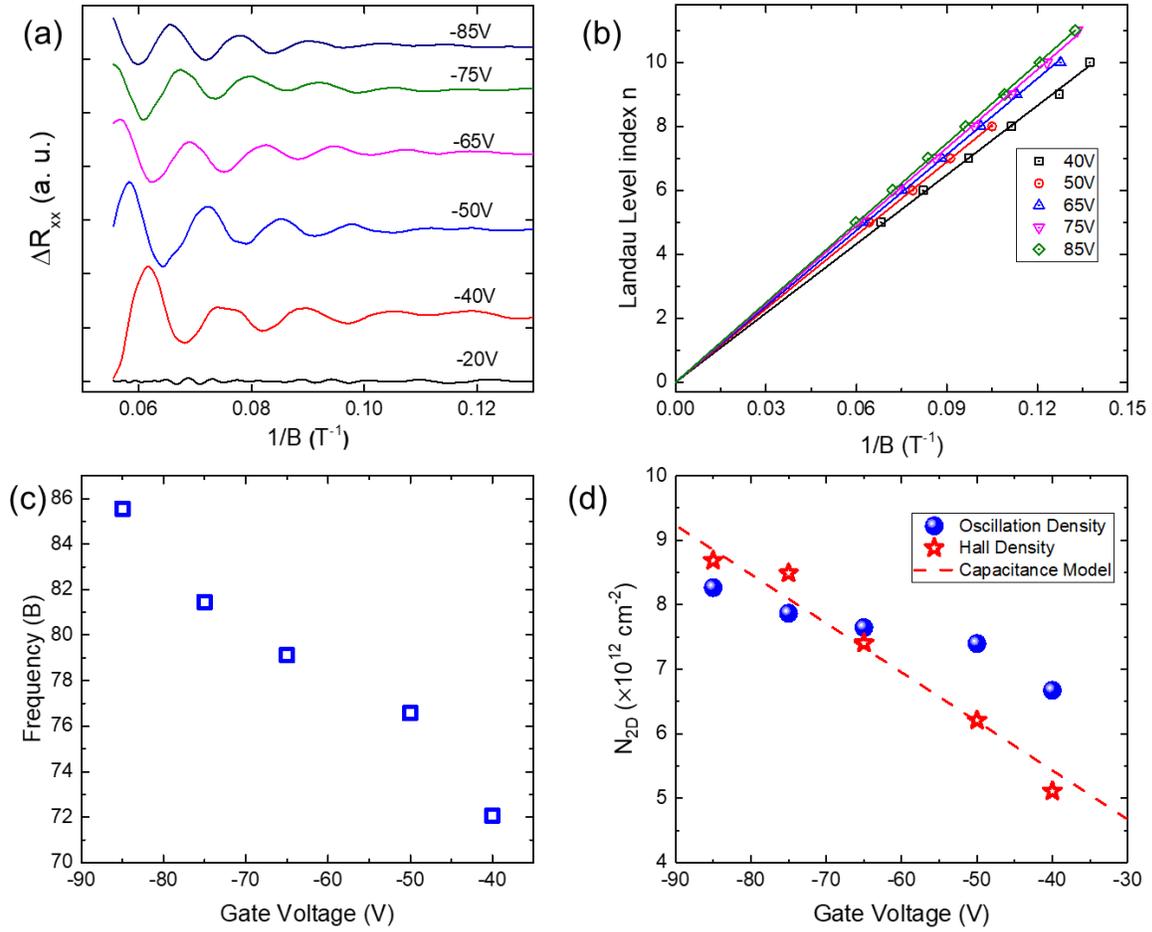

**Figure 2 Gate-dependence of Shubnikov-de-Haas oscillations versus 1/B. (a)** The oscillation amplitude under different gate biases. Oscillations were revolved only when the negative gate bias is large enough. The amplitudes were extracted by subtracting a smooth background from $R_{xx}$. **(b)** Landau fan diagram under different gate bias. Each data point represents a minimum in oscillation amplitudes. All the lines have intercept near zero, indicating standard phase of oscillations. **(c)** Oscillation frequencies as a function of gate bias. The frequencies were extracted from the slope in (b). **(d)** Comparison of 2D carrier density measured from oscillation frequencies (pink sphere), Hall resistance (orange pentagrams) and calculated from capacitance model (dotted line).

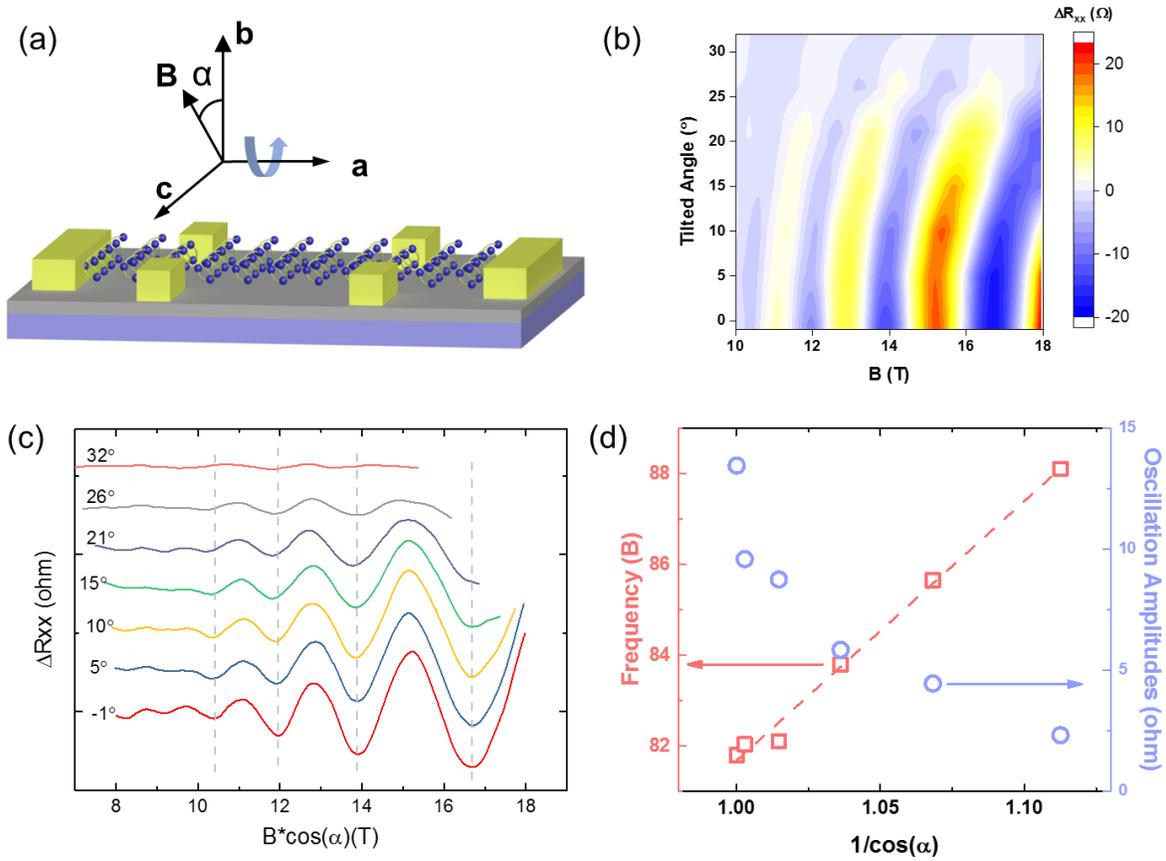

**Figure 3 Angle-dependence of oscillation frequencies.** (**a**) Schematic configuration for angle-resolved oscillation frequency measurement. (**b**) The color map of oscillation amplitude as a function of B field and tilted angles. (**c**) Oscillation amplitudes plotted against perpendicular magnetic field component B×cos(α). (**d**) Oscillation frequencies(red squares, left axis) and amplitudes (blue circles, right axis) as a function of 1/cos (α). The dash line shows a linear relationship between oscillation frequency and 1/cos(α), a signature of 2D Fermi surface.

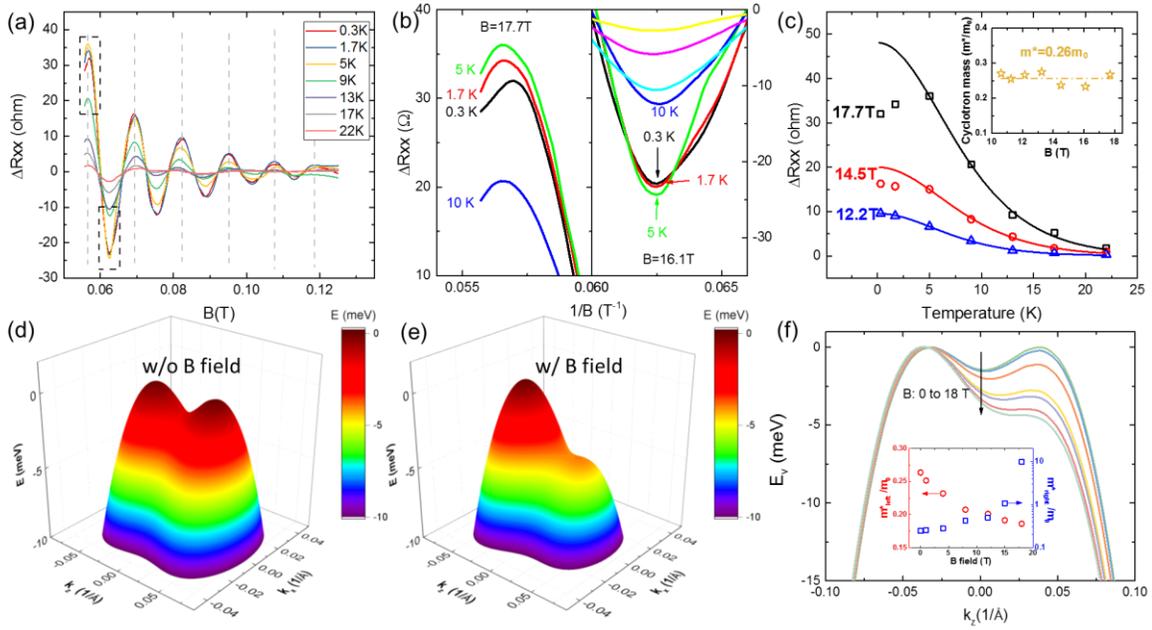

**Figure 4 (a)** Temperature-dependent oscillation amplitudes. T **(b)** Zoomed-in features of two oscillation peaks at 17.7 T and 16.1 T as highlighted in dash boxes in (a). **(c)** Extracted oscillation peak amplitudes under different magnetic field as a function of temperature. The solid line is the fitting curve from Lifshitz-Kosevich equation with data points T>5K. Under large magnetic field, the low-temperature data points deviated from fitting curves, which was also captured in (b). **Inset**: the effective cyclotron mass was extracted to be 0.26 $m_0$. **(d)** and **(e)**: the camel-back shape of valence band maxima vicinity w/o and w external B field. **(f)** A cross-section of valence band edge cut along $k_z$ direction. **Inset**: the effective mass of light and heavy pocket holes under different magnetic fields.

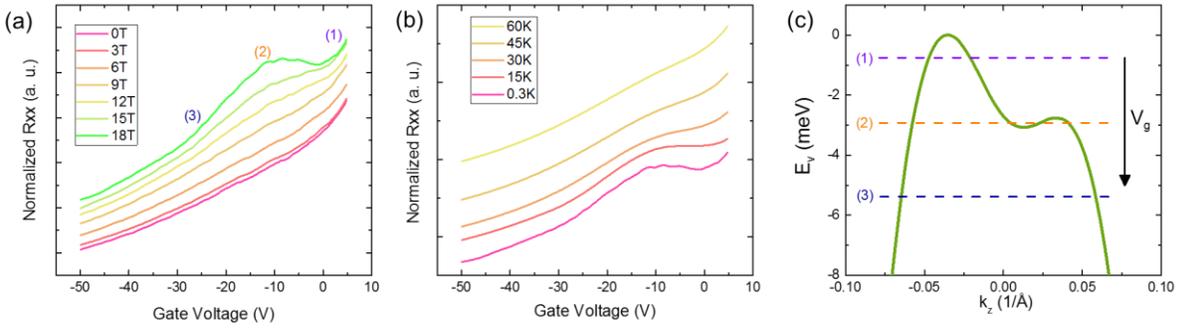

**Figure 5** Anomalous increase in Rxx when sweeping gate voltage towards negative bias under **(a)** different magnetic field and **(b)** different temperature. **(c)** Explanation of the abnormal increase of resistance. As we sweep gate in reverse direction, the Fermi energy will be pulled downwards and slightly enter the valence band. (1), (2) and (3) represent three typical scenarios corresponding to the points with the same notations in (a).

**Acknowledgement**   P. D. Y. would like to thank W. Pan, T. Low and Y. Lyanda-Geller for the valuable discussions. P. D. Y. was supported by NSF/AFOSR 2DARE Program and SRC GRC Program. W. Z. W. was partially supported by a grant from the Oak Ridge Associated Universities (ORAU) Junior Faculty Enhancement Award Program. Part of the solution synthesis work was supported by the National Science Foundation under grant no. CMMI-1663214. P. D. Y. and W. Z. W. were also support by Army Research Office under grant no. W911NF-15-1-0574 and W911NF-17-1-0573. P. D. Y. and K. C. were also supported in part by ASCENT, one of six centers in JUMP, a SRC program sponsored by DARPA. A portion of this work was performed at the National High Magnetic Field Laboratory, which is supported by National Science Foundation Cooperative Agreement No. DMR-1644779 and the State of Florida.

**Author Contributions** P. D. Y. conceived and supervised the project. P. D. Y. and G. Q. designed the experiments. Y. W. and W. Z. W. synthesized the material.  G. Q. fabricated the devices and performed the low-temperature magneto-transport measurement. P. D. Y. and G. Q. analyzed the data. Y. N., Y. Z. and K. C. performed DFT calculations. G. Q. and P. D. Y. wrote the manuscript. All authors have discussed the results and commented on the paper.

**Competing financial interest statement**   The authors declare no competing financial interests.

Supplementary Materials for:

# Quantum Transport and Band Structure Evolution under High Magnetic Field in Few-layer Tellurene

Gang Qiu[1,2], Yixiu Wang[3], Yifan Nie[4], Yongping Zheng[4], Kyeongjae Cho[4], Wenzhuo Wu[2,3],

Peide D. Ye[1,2]*,

[1]School of Electrical and Computer Engineering, Purdue University, West Lafayette, Indiana 47907, USA

[2]Birck Nanotechnology Center, Purdue University, West Lafayette, Indiana 47907, USA

[3]School of Industrial Engineering, Purdue University, West Lafayette, Indiana 47907, USA

[4]Department of Materials Science and Engineering, the University of Texas at Dallas, Richardson, Texas 75080, United States

*e-mail: Correspondence and requests for materials should be addressed to P. D. Y. (yep@purdue.edu)

**Supplementary Note 1: Synthesis Method of Large-area Tellurene Nano-films**

The synthesis of tellurene nano-films follows the procedure described in the previous literature[1]. 0.09 g of $Na_2TeO_3$ and 0.50 g of poly (vinyl pyrrolidone) (PVP, M.W.= 58000) were put into a 50 mL Teflon-lined stainless steel autoclave. Subsequently aqueous ammonia solution (25–28%, w/w%) and hydrazine hydrate (80%, w/w%) with the volume ratio of 2:1 were added into the mixed solution under vigorous magnetic stirring. Double-distilled water was added to 80% of the container volume. The container was then sealed and heated at 180 °C for 20 hours before cooling down to room temperature.

The freshly prepared solution with 2D Te suspension was centrifuged. After removing the supernate, N, N-dimethylformamide (DMF; 1 mL) and $CHCl_3$ (1 mL) were added into the mixture. A 50 μL syringe was used to disperse mixed suspension containing Te flakes onto the water drop by drop. After that, the floating 2D Te can be directly scooped on the substrate.

**Supplementary Note 2: Device Fabrication and Characterization**

Figure S1 shows the fabrication process flow for Hall bar devices. The as-grown tellurene flakes are usually in rectangular or trapezoidal shapes, with one-dimensional atomic chains ([0001] crystal orientation) aligned with the long edge of the flakes, which was studied thoroughly in previous Raman experiments[2]. After being dispersed onto 300 nm $SiO_2$/p++ Si substrates, the samples were cleaned by standard solvent cleaning procedure and patterned with electron beam lithography into two rectangles with identical geometry but along different primary crystal axes. The flake is then trimmed with 30 seconds of $BCl_3$/Ar plasma dry etching process. The purpose of this step is to eliminate the geometric non-ideality in measuring the transport anisotropy. Six-terminal Hall-bar structure was subsequently patterned by electron beam lithography again and metal contacts of 50/100 nm Pd/Au were formed by electron beam evaporator. We denote the bottom flake in Figure S1(c) as along 1D direction since the drain current flows along 1D atom chains and the top flake as vdW direction since the drain current flows along van der Waals bonds.

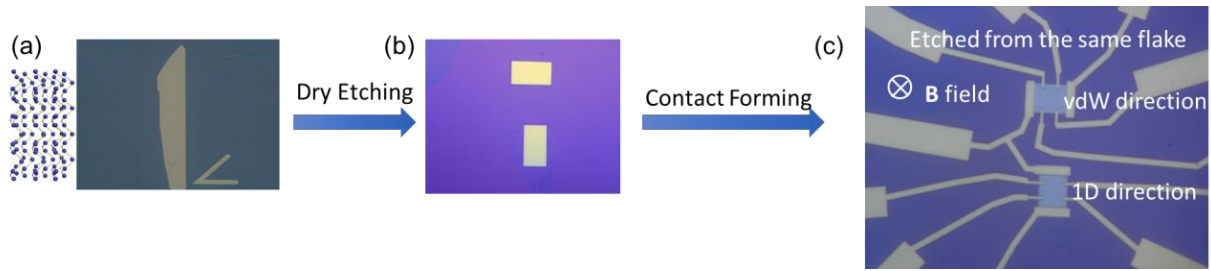

**Figure S1. Process flow of fabricating Hall bar devices along different directions.** The optical image of **(a)** as-grown tellurene flakes dispersed on the substrates. The 1D atomic chains are aligned with the long edge of the flake; **(b)** the flakes were trimmed into to perpendicular rectangles with BCl$_3$/Ar dry etching method; **(C)** Pd/Au contacts were patterned and metallized to form two Hall-bar devices.

The magneto transport measurement was carried out in a He$^3$ cryostat with a superconducting magnet at National High Magnetic Field Lab in Tallahassee, FL. The data was acquired from Stanford Research Instruments SR830 lock-in amplifier with standard lock-in measurement techniques.

**Supplementary Note 3: Density functional theory (DFT) calculations**

The density functional theory (DFT) calculations are performed in two parts. The structural optimization of the crystal structure is performed with the Vienna ab-initio Simulation Packages (VASP)[3], with the projector-augmented wave (PAW) method[4]. The wave functions are expanded into plane waves with an energy cut-off of 400 eV. Exchange-correlation interactions are described by the generalized gradient approximation (GGA) using the Perdew-Burke-Ernzerhof (PBE) functional[5]. A Γ-centered 8x8x8 k-point sampling mesh is adopted in the integration in the first Brillouin zone. The convergence criteria for electron and ionic minimization are $10^{-4}$ eV and $10^{-2}$ eV/Å, respectively. For a better description of the electronic structures, the band structure calculation is performed with WIEN2K[6], using the modified Becke-Johnson local-density approximation (MBJLDA) exchange functionals[7]. Spin-orbit coupling is implemented in the calculation, with a Γ-centered 12x12x12 k-point mesh. To calculate the electronic band structure change under external magnetic field, the spin up and spin down exchange correlation potentials are shifted by $\pm\mu_B B_{ext}$, respectively[8]. The external magnetic field is normal to the helix axis.

**Supplementary Note 4: Te band structure from DFT calculations**

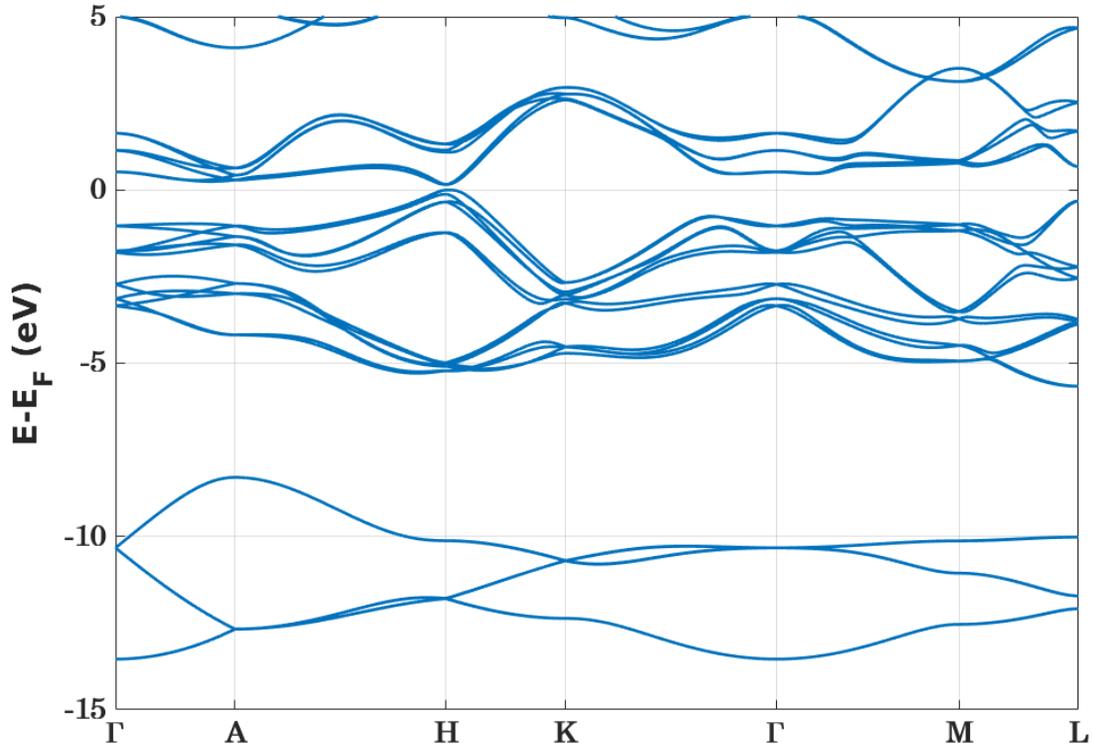

**Figure S2.** Full electronic structure of bulk Te calculated from DFT methods as discussed in Note 3.

**Supplementary Note 5: Hall Measurement along 1D Direction and van der Waals Direction**

The conductance (1/$R_{xx}$) along 1D direction and vdW direction was first measured under different gate bias under low temperature. The 1D direction shows about 1.3 times higher conductivity than vdW direction. Since the conductance can be expressed as $\sigma = \frac{1}{\rho_{xx}} = qn\mu$, which is proportional to the carrier density and mobility, one may intuitively reckon that the carrier density should be the same along two directions whereas the conductance discrepancy should arise from the mobility difference. We attempted to verify this presumption by conducting Hall measurement on both devices and calculate 2D carrier density $n_{2D}$ and Hall mobility $\mu_{Hall}$ from the equations: $n_{2D} = \frac{B}{eR_{xy}}$ and $\mu_{Hall} = \frac{L}{W}\frac{1}{R_{xx}n_{2D}e}$ respectively. However, the calculated Hall mobility is similar in both devices yet the 2D carrier density varies with a ratio similar to the conductance anisotropic ratio, as shown in Figure S3(b), which conflicts with our anticipation. In order to correctly evaluate the results from the Hall measurement, we have to retrospect the derivation of above equations in calculating 2D sheet density and Hall mobility. The root of Hall effect is the balance between Lorentz force and Coulomb force: $q\boldsymbol{v} \times \boldsymbol{B} = q\boldsymbol{E}$, where the carrier velocity is generalized with the average velocity. However, in real scenarios the distribution of velocity of electrons should follow Boltzmann distribution. For those electrons travelling faster than average velocity, the Lorentz force will be larger than Coulomb force and the trajectory of the particle will be deviated in one direction, and likewise the electrons with smaller velocity will be deviated into the other direction, as depicted in Figure S4. In other words, the Hall mobility measures average in-plane motion, so it is reasonable for both devices to have similar Hall mobility. For 2D carrier densities, if we follow the classical Boltzmann distribution and take all electrons with different velocity into account by integrating by velocity, than we shall alter the widely used formula $n_{2D} = \frac{B}{eR_{xy}}$ into a slight different form $n_{2D} = r_H \frac{B}{eR_{xy}}$, here $r_H$ is the Hall factor which is defined as: $r_H = \frac{\langle \tau_m^2 \rangle}{\langle \tau_m \rangle^2}$. Since the relaxation time $\tau_m$ is a tensor, $r_H$ contains the information of anisotropic transport. Hence, it is questionable to use Hall measurement to calibrate anisotropic transport and alternative approaches should be adopted as we will discuss in the next session.

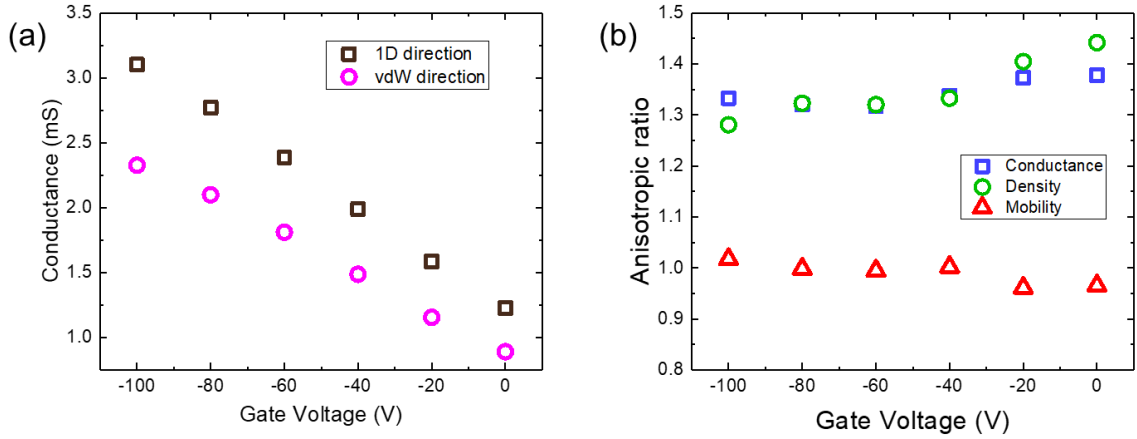

**Figure S3.** (a) Conductance measured under different gate bias. (b) Anisotropic ratio of conductance, 2D carrier density and Hall mobility. The 2D carrier density $n_{2D}$ and Hall mobility $\mu_{Hall}$ from the equation: $n_{2D} = \frac{B}{eR_{xy}}$ and $\mu_{Hall} = \frac{L}{W}\frac{1}{R_{xx}n_{2D}e}$ respectively.

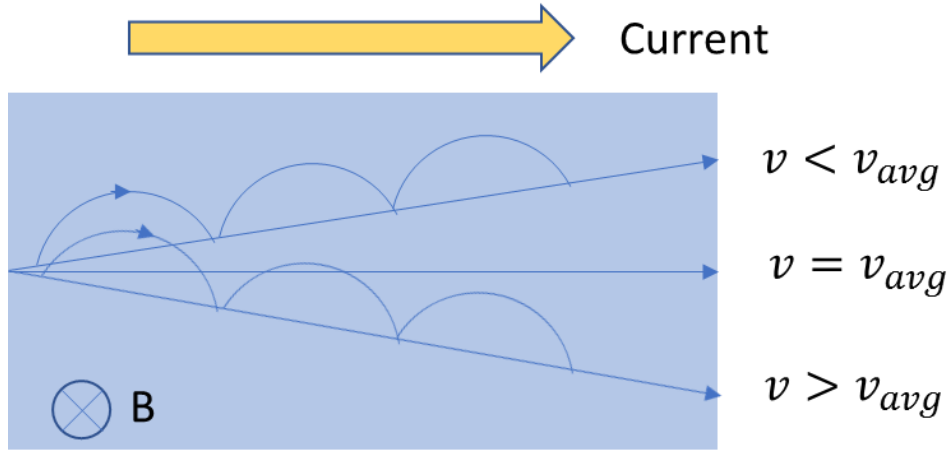

**Figure S4.** Schematic demonstration of velocity distribution in Hall measurement.

**Supplementary Note 6: Determination of Mobility Anisotropic Ratio from Magneto-Resistance**

Another common practice of determining mobility is to use magneto-resistance measurement. In small magnetic field regime, the magneto-resistance (defined as $\Delta MR = \frac{\rho_B - \rho_0}{\rho_0}$) should follow a parabolic relationship with the external magnetic field: $\Delta MR = \alpha\mu^2 B^2$, where $\alpha$ is the magneto resistance coefficient that is only related with geometric factor. Hence for two identical devices, we can assume the coefficient $\alpha$ to be the same and hence by fitting the magneto resistance in B<5T regime, we can derive the mobility ratio to be $\mu_{1D}/\mu_{vdW} = 1.32$. This value is close to room temperature field-effect mobility ratio and was explained by DFT calculations.

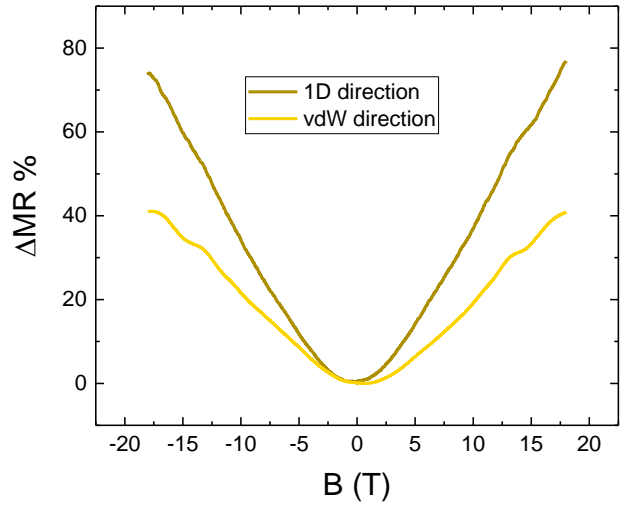

**Figure S5.** Magneto resistance along 1D direction and vdW direction. At small magnetic field regime (B < 5 T), the magneto resistance follows parabolic relationship with B field, from which we can extract the mobility.

**Supplementary Note 7: SdH Oscillations in thicker samples**

We found that in another thicker sample (18 nm), the amplitude of oscillations are much weaker and only two periods can be observed. And no plateau is formed in transverse resistance, indicating much weaker quantum confinement in thicker tellurene films.

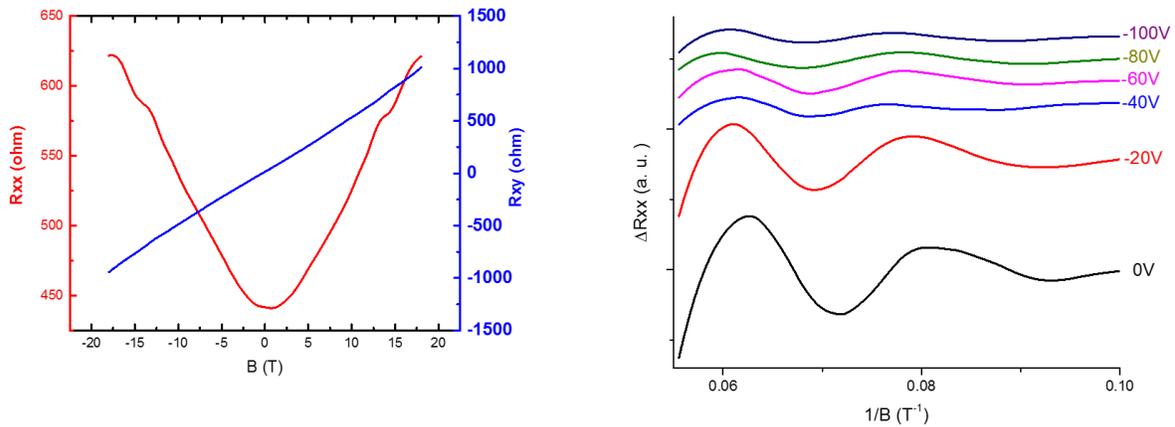

**Figure S6.** Shubnikov-de-Haas oscillations in an 18-nm-thick sample. The oscillation features are less predominant compared to the 10-nm-thick sample, indicating weaker quantum confinement in thicker tellurene films.